\documentstyle[12pt]{article}
\catcode`\@=11
\def\@citex[#1]#2{\if@filesw\immediate\write\@auxout
	{\string\citation{#2}}\fi
\def\@citea{}\@cite{\@for\@citeb:=#2\do
	{\@citea\def\@citea{,}\@ifundefined
	{b@\@citeb}{{\bf ?}\@warning
	{Citation `\@citeb' on page \thepage \space undefined}}
	{\csname b@\@citeb\endcsname}}}{#1}}

\newif\if@cghi
\def\cite{\@cghitrue\@ifnextchar [{\@tempswatrue
	\@citex}{\@tempswafalse\@citex[]}}
\def\citelow{\@cghifalse\@ifnextchar [{\@tempswatrue
	\@citex}{\@tempswafalse\@citex[]}}
\def\@cite#1#2{{$\null^{#1)}$\if@tempswa\typeout
	{warning: optional citation argument
	ignored: `#2'} \fi}}

\begin{document}
\thispagestyle{empty}

\begin{titlepage}

\centerline{\normalsize hep-ph/9506244 \hfill OITS-577}
\centerline{\normalsize \hfill May 1995}
\mbox{}

\vspace{4cm}

\begin{center}

{\bf STATUS OF PERTURBATIVE QCD EVALUATION OF HADRONIC DECAY RATES
     OF THE Z AND HIGGS BOSONS}

\vspace{1.3cm}

     {\bf Davison E.\ Soper} and {\bf Levan R.\ Surguladze}

{\it Institute of Theoretical Science, University of Oregon,\\
                  Eugene, OR 97403, USA}

\vspace{3mm}

             (presented by L.\ R.\ Surguladze)
\end{center}

\vspace{6.3cm}

{\small
  We review the current status of the high order perturbative
QCD evaluation of the hadronic decay rates of the Z and Higgs bosons.
A systematic classification of the various types of QCD corrections to
$O(\alpha_s^2)$ and $O(\alpha_s^3)$ is made and their numerical
status is clarified.}

\vfill

\centerline{\it to appear in the Proceedings of the XXXth Rencontres
de Moriond}
\centerline{\it ``QCD and High Energy Hadronic Interactions''}
\centerline{\it Les Arcs, France, March 19-26, 1995}

\end{titlepage}

\section{Introduction}
The standard theory of strong interactions, QCD, can be
acurately tested based on the precise experiments at LEP/SLC and the
recent high order perturbative results for the hadronic width of the
Z boson. From the fit of theoretical and experimental results,
an acurate value of the strong coupling can be extracted\cite{REV}.
With the planned
experimental precision at LEP, an error of a couple of percent
on $\alpha_s$ is expected\cite{LEP}.
 In this situation, a further improvement of the
theoretical precision for the measured quantities is necessary.
In this work, we present a systematic classification of various types of
high order QCD contributions to the hadronic decay rates of the Z boson.
We also discuss recent results of the evaluation of the $O(\alpha_s^2)$
QCD corrections to the hadronic decay rates of the Standard Model Higgs
boson.

\section{Hadronic decay width of the Z boson}

\subsection{Theoretical structure}

The decay rate of the Z boson into quark antiquark pair can be written
in the following way

\begin{eqnarray}
\lefteqn{\hspace{-20mm}\Gamma_{Z\rightarrow \mbox{\scriptsize hadrons}}
        =\frac{G_FM_Z^3}{8\sqrt{2}\pi} \sum_f \rho_f\biggl\{
          v_f^2\biggl[(1+2X_f)\sqrt{1-4X_f}
          +\delta_{\mbox{\tiny QCD}}^{\mbox{\tiny V}}(\alpha_s,X_f,X_t)
          +\delta_{\mbox{\tiny QED}}^{\mbox{\tiny V}}(\alpha,\alpha_s,X_f)
                                         \biggr]}\nonumber\\
 && \quad \hspace{15mm}
          +a_f^2\biggl[(1-4X_f)^{3/2}
          +\delta_{\mbox{\tiny QCD}}^{\mbox{\tiny A}}(\alpha_s,X_f,X_t)
          +\delta_{\mbox{\tiny QED}}^{\mbox{\tiny A}}(\alpha,\alpha_s,X_f)
                                                        \biggr]
                                                             \biggr\},
\label{Zqq}
\end{eqnarray}
where $v_f=2I_f^{(3)}-4e_f\sin^2\theta_{\mbox{\tiny W}}k_f$,
$a_f=2I_f^{(3)}$ are the standard vector and axial couplings.
$X_f=m_f^2/M_Z^2$, e.g., $X_b \sim 0.003$.
$\delta_{\mbox{\tiny QCD}}^{\mbox{\tiny V/A}}$
and $\delta_{\mbox{\tiny QED}}^{\mbox{\tiny V/A}}$ stand for the
vector and axial parts of the corresponding QCD and QED contributions.
The summation $f=u,d,s,c,b$ should be done with a proper care for the
singlet contributions, which are included in
$\delta_{\mbox{\tiny QCD}}^{\mbox{\tiny V/A}}$ (see below).
The electroweak self-energy and vertex corrections are absorbed in
the factors $\rho_f$ and $k_f$. The current status of the
electroweak contributions has been discussed in detail\cite{Knirev}.
The small QED corrections in vector and axial channels look like
\begin{equation}
\delta_{\mbox{\tiny QED}}^{\mbox{\tiny V}}
     =\frac{3}{4}e_f^2\frac{\alpha}{\pi}[1+12X_f+O(X_f^2)]
                       +O(\alpha^2)+O(\alpha\alpha_s),
\label{QEDV}
\end{equation}
\begin{equation}
\delta_{\mbox{\tiny QED}}^{\mbox{\tiny A}}
     =\frac{3}{4}e_f^2\frac{\alpha}{\pi}[1-6X_f
                   -12X_f\log X_f+O(X_f^2)]
                       +O(\alpha^2)+O(\alpha\alpha_s).
\label{QEDA}
\end{equation}
Corrections $\sim \alpha^2$ and $\sim \alpha\alpha_s$ are
negligible\cite{katQED} at the current level of precision.

\subsection{QCD contributions to $O(\alpha_s^3)$}

The QCD contributions are represented by the terms
$\delta_{\mbox{\tiny QCD}}^{\mbox{\tiny V/A}}$ and can be  calculated
within perturbation theory by evaluating the quantity
Im$\Pi(-s+i0)$ at $s=M_Z^2$,
where the function $\Pi$ is defined through a correlation function
of two flavor diagonal quark currents
\begin{equation}
 i\int d^4x e^{iqx}<Tj^f_{\mu}(x)j^f_{\nu}(0)>_{0}
   =g_{\mu\nu}\Pi(Q^2)-Q_{\mu}Q_{\nu}\Pi'(Q^2).
\label{PI}
\end{equation}
Here, $Q^2$ is a large ($\sim -M_Z^2$) Euclidean momentum
and the neutral weak current of quark coupled to Z boson is
\begin{equation}
j_{\mu}^{f}=\biggl(\frac{G_FM_Z^2}{2\sqrt{2}}\biggr)^{1/2}
      (v_{f}\overline{q}_f\gamma_{\mu}q_f
      +a_{f}\overline{q}_f\gamma_{\mu}\gamma_5q_f).
\label{j}
\end{equation}
Because the physical scale ($\sim M_Z$) is much larger than the
quark masses involved, we expand the $\Pi$ function in powers of
$m_f^2/M_Z^2$.
The coefficient functions in this expansion and their imaginary parts can be
calculated using the methods of the renormalization group\cite{RG}
and the computer programs for analytical evaluation
of multiloop Feynman diagrams\cite{CompRev}. For a review of the current
state-of-art of the high order perturbative QCD calculations for the
above quantities see the Ref.\ 7.

  It is convenient to decompose
$\delta_{\mbox{\tiny QCD}}^{\mbox{\tiny V/A}}$ in so called singlet and
non-singlet parts. The nonsinglet part is formed from the diagrams
with the two quark currents within a single fermionic loop and the
singlet part corresponds to the graphs with the quark currents in separate
fermionic loops mediated by gluonic states.
$
\delta_{\mbox{\tiny QCD}}^{\mbox{\tiny V}}
=\delta_{\mbox{\tiny QCD}}^{\mbox{\tiny V,\scriptsize ns}}
+\delta_{\mbox{\tiny QCD}}^{\mbox{\tiny V,\scriptsize s}}/v_f^2,
\hspace{2mm}
\delta_{\mbox{\tiny QCD}}^{\mbox{\tiny A}}
=\delta_{\mbox{\tiny QCD}}^{\mbox{\tiny A,\scriptsize ns}}
+\delta_{\mbox{\tiny QCD}}^{\mbox{\tiny A,\scriptsize s}}/a_f^2.
$

The nonsinglet part in the vector channel to $O(\alpha_s^3)$ reads
\begin{eqnarray}
\lefteqn{\hspace{-20mm}\delta_{\mbox{\tiny QCD}}^{\mbox{\tiny V,\scriptsize
ns}}
=\frac{\alpha_s}{\pi}(1+12\overline{X}_f)
+\biggl(\frac{\alpha_s}{\pi}\biggr)^2(1.40923+104.833\overline{X}_f
+\sum_v F^{(2)}(X_v) + G^{(2)}(X_t))}\nonumber\\
 && \quad \hspace{9mm}
+\biggl(\frac{\alpha_s}{\pi}\biggr)^3(-12.76706+547.879\overline{X}_f
     +\sum_v F^{(3)}(X_v) + G^{(3)}(X_t)),
\label{Vns}
\end{eqnarray}
where $\overline{X}_f=m_f^2(M_Z)/M_Z^2$.
The $O(\alpha_s^2)$ and $O(\alpha_s^3)$ terms in the limit of vanishing
light quark masses $\overline{X}_f$=0 and infinitely large top mass
$m_t=\infty$ have been evaluated\cite{Dine,Rs}.
The terms $\sim \overline{X}_f$
represent leading mass corrections\cite{Chetm2V,Sop,Zhm}.
The functions $F^{(2)}(X_f)$ and $G^{(2)}(X_t)$ stand for the contributions
from the three-loop diagrams containing
the virtual fermionic loop with the massive quark propagating in it.
As we will see, these
contributions are already small, so there is no necesity to evaluate
light quark mass corrections for $F$ and $G$.
 The function $F$ represents the
contribution of light (first five flavors) quarks, while the function
$G$ represents the remaining contribution of the decoupled top quark in
five flavor effective theory. Numerically
\cite{Sop},
\begin{equation}
F^{(2)}(X_v) \approx \overline{X}_v^2 \times \left\{
-0.474894
- \log \overline{X}_v
+\sqrt{\overline{X}_v}  \left[
-0.5324
+0.0185 \log \overline{X}_v \right]
\right\}
\end{equation}
is small and can be neglected.
Contributions from virtual top quark involve
$G(X_t)$ function. We see that $G$ vanishes in the large $X_t$ limit.
That is, the heavy quark decouples\cite{Sop}
\begin{equation}
G^{(2)}(X_t) \approx { \overline{X}_t^{-1}}\times\left\{
  {44 \over 675}
+ {2 \over 135} \log \overline{X}_t
-\sqrt{\overline{X}_t^{-1}}
\left[0.001226
+0.001129\log \overline{X}_t
\right]
\right\}.
\label{Glargem}
\end{equation}
The three-loop analytical expressions for the $F$ and $G$ functions
were found in
Ref.\ 13. Also, the first two terms in the r.h.s. of eq.(\ref{Glargem})
have been obtained using the large mass expansion methods\cite{Chetvirt}.
At order $\alpha_s^3$,
$F^{(3)}(X_f)=-6.12623 \overline{X}_f+O(\overline{X}_f^2)$ \cite{Zhm}.
Based on the large mass expansion technique, the following tiny correction
has been obtained for the virtual top quark contribution
in the limit $X_t \rightarrow \infty$ \cite{Lar}
\begin{equation}
G^{(3)}(X_t) \approx { \overline{X}_t^{-1}}\times [
-0.1737-0.2124\log \overline{X}_t - 0.0372 \log^2\overline{X}_t]
\label{Glargem3}
\end{equation}
In these formulas,
$\alpha_s$ denotes running $\overline{\mbox{MS}}$ coupling in five
flavor theory evaluated at $M_Z$. The transformation relation for different
number of flavors and different scales, as well as the relation between the
$\overline{\mbox{MS}}$ running mass and the pole mass can be found in
Ref.\ 16.

The nonsinglet part in the axial channel is very similar to the one in
the vector channel, except the light quark mass corrections are different.
\begin{eqnarray}
\lefteqn{\hspace{-16mm}\delta_{\mbox{\tiny QCD}}^{\mbox{\tiny A,\scriptsize
ns}}
=\frac{\alpha_s}{\pi}(1-22\overline{X}_f)
+\biggl(\frac{\alpha_s}{\pi}\biggr)^2(1.40923-85.7136\overline{X}_f
+\sum_v F^{(2)}(X_v) + G^{(2)}(X_t))}\nonumber\\
 && \quad \hspace{13mm}
+\biggl(\frac{\alpha_s}{\pi}\biggr)^3(-12.76706+(\mbox{\small unknown})
                                                            \overline{X}_f
     +\sum_v F^{(3)}(X_v) + G^{(3)}(X_t)).
\label{Ans}
\end{eqnarray}
The terms $\sim \overline{X}_f$ represent leading light quark mass
corrections\cite{Chetm2A,Sop,Zhm}.

    In the vector channel,
the singlet $O(\alpha_s^2)$ contribution is identically zero due to Furry's
theorem\cite{Furry}. In the axial channel, at the same order,
the light doublet contributions add up to zero in the limit of
degenerate quark masses.
This is because in the Standard Model, quarks in a weak doublet couple
with opposite sign to the Z boson in the axial channel.
However, the contribution from the t,b doublet turnes out to
be significant due to the large mass splitting\cite{Knitrian}
\begin{eqnarray}
\lefteqn{\hspace{-14mm}{\cal L}_{\mbox{\scriptsize A}}^{(2)}=
\biggl(\frac{\alpha_s}{\pi}\biggr)^2\biggl[
\biggl(-\frac{37}{12}-\log X_t+\frac{7}{81}X_t^{-1}
                                 +0.013X_t^{-2}\biggr)_{m_b=0} }\nonumber\\
 && \quad
+\overline{X}_b(18+6\log X_t)
-\frac{\overline{X}_b}{X_t}\biggl(\frac{80}{81}+\frac{5}{27}\log X_t\biggr)
\biggr],
 \label{Axtri}
\end{eqnarray}
where small mass corrections $\sim \overline{X}_b$ have been calculated
in Ref.\ 20. At the $O(\alpha_s^3)$, both channels contribute.
The vector channel contribution in the limit of massles light
quarks reads\cite{Rs}
\begin{equation}
{\cal L}_{\mbox{\scriptsize V}}^{(3)}=\biggl(\frac{\alpha_s}{\pi}\biggr)^3
\biggl[-0.41318(\sum_{f=u,d,s,c,b}v_f)^2+
(0.02703\overline{X}_t^{-1}+0.00364\overline{X}_t^{-2}+O(X_t^{-3}))
v_t\sum_{f=u,d,s,c,b}v_f\biggr],
\label{Vtri3}
\end{equation}
where negligible terms $\sim \overline{X}_t^{-1}, \overline{X}_t^{-2}$
were computed in Ref.\ 15.
In the axial channel, the $O(\alpha_s^3)$ singlet contribution in the
large top mass expansion reads\cite{Chettri3,Lar,Knirev}
\begin{equation}
{\cal L}_{\mbox{\scriptsize A}}^{(3)}=\biggl(\frac{\alpha_s}{\pi}\biggr)^3
\biggl(-15.98773 -\frac{67}{18}\log X_t + \frac{23}{12}\log ^2X_t\biggr).
\label{Atri3}
\end{equation}
The light quark mass corrections at the $O(\alpha_s^3)$ for the singlet
parts are not yet known. However, at the current experimental precision
they are not expected to be detectable. Summarizing the present knowledge
for the singlet parts, we write
$\delta_{\mbox{\tiny QCD}}^{\mbox{\tiny V,\scriptsize s}}
={\cal L}_{\mbox{\scriptsize V}}^{(3)}$, \hspace{3mm}
$\delta_{\mbox{\tiny QCD}}^{\mbox{\tiny A,\scriptsize s}}
={\cal L}_{\mbox{\scriptsize A}}^{(2)}+{\cal L}_{\mbox{\scriptsize A}}^{(3)}$.

\subsection{On large ``$\pi^2$'' terms}

Let us briefly mention about the importance of the so called $\pi^2$ terms
appearing naturally as a result of analytical continuation of the results
of perturbative evaluation from Euclidean to Minkowski space. Indeed,
because of the relation $\frac{1}{\pi}$Im $\log^3(s+i0)=-3\log^2s+\pi^2$
and relations similar to that, one can trace the appearance of the
large contributions due to the $\pi^2$ terms starting at $O(\alpha_s^3)$.
For the known quantity $R(s)$ in electron-positron annihilation at
low energies, we write
\begin{eqnarray}
\lefteqn{\hspace{-18mm}R=1+\frac{\alpha_s}{\pi}d_1
+\biggl(\frac{\alpha_s}{\pi}\biggr)^2d_2
+\biggl(\frac{\alpha_s}{\pi}\biggr)^3(d_3-\pi^2\beta_0^2d_1/3)
+\biggl(\frac{\alpha_s}{\pi}\biggr)^4[(d_4
-\pi^2(\beta_0^2d_2+5\beta_0\beta_1d_1/6)]}\nonumber\\
 && \quad \hspace{-3mm}
+\biggl(\frac{\alpha_s}{\pi}\biggr)^5\{d_5
-\pi^2[2\beta_0^2d_3+7\beta_0\beta_1d_2/3
+(\beta_0\beta_2+\beta_1^2/2)d_1]+\pi^4\beta_0^4d_1/5\}.
\label{pi^2}
\end{eqnarray}
The $\beta$ function coefficients can be found in, e.g., Ref.\ 16.
The $d_i$ are known up to $i=3$ and are of order 1
($d_i=\{1,1.4,-0.7\}$)\cite{Rs}.
The $\pi^2$ terms are large. They can be calculated up to order $\alpha_s^5$
using the known $d_i$ and $\beta_i$. Numerically, they are
$\{-12.1(\alpha_s/\pi)^3,-89.2(\alpha_s/\pi)^4,-648(\alpha_s/\pi)^5\}$.
It is reasonable to expect that the contributions $\sim\pi^2$
are dominant at all orders and their resummation seems to be of
primary importance. From the above equation, one can also
get the idea about the size of the still uncalculated higher order
corrections and numerical error estimates.

\section{Hadronic decay width of the Higgs boson}

The perturbative QCD evaluation of the hadronic decay rates of the SM Higgs
boson\cite{Knirev,Gra}
is a very similar to that of the Z boson. In fact, exactly the same set of
diagrams have to be evaluated at each order. However, in this case one
considers quark scalar current correlators. Also, the calculation involves
renormalization of quark mass, even in the massless quark limit.
This is because of the explicit quark mass dependence of the Yukawa coupling.
We give the expression for the quantity
$\Gamma_{H\rightarrow \mbox{\scriptsize hadrons}}$
to $O(\alpha_s^2)$,
without detailing the particular contributions from the singlet and
nonsinglet parts.
\begin{eqnarray}
\lefteqn{\Gamma_{H\rightarrow \mbox{\scriptsize hadrons}}
        =\frac{3\sqrt{2}G_FM_H}{8\pi}\sum_{f}\overline{m}_f^2
\biggl(1+\frac{\alpha}{\pi}e_f^2\Delta_{\mbox{\scriptsize em}}\biggr)
\times(1+\Delta_{\mbox{\scriptsize weak}})
\times\biggl[\biggl(1-4\frac{\overline{m}_f^2}{M_H^2}\biggr)^{3/2} }\nonumber\\
 &&
+\frac{\alpha_s}{\pi}\biggl(5.667-40\frac{\overline{m}_f^2}{M_H^2}\biggr)
+\biggl(\frac{\alpha_s}{\pi}\biggr)^2\biggl(29.147
                       -99.725\frac{\overline{m}_f^2}{M_H^2}
+12\sum_{v=u,d,s,c,b}\frac{\overline{m}_v^2}{M_H^2}\biggr),
\label{Hqq}
\end{eqnarray}
where $\overline{m}_f$ is the $\overline{\mbox{MS}}$ quark mass evaluated
at $M_H$.
The leading electroweak contributions, represented by the terms
$\Delta_{\mbox{\scriptsize em}}$ and $\Delta_{\mbox{\scriptsize weak}}$,
are calculated in Refs.\ 23,24,3. The leading QCD corrections
$\sim \alpha_s G_Fm_t^2$ have been calculated in Ref.\ 25.
The $O(\alpha_s)$ QCD correction with the exact quark mass dependence
was found in Ref.\ 26. The $O(\alpha_s^2)$ corrections
for the scalar quark current correlator and to the Higgs decay rates have been
calculated in Refs.\ 27,16. The last term in the above expression is
due to nonvanishing quark masses from the nonsinglet
diagrams containing virtual fermionic loop and the singlet diagrams.
Here we assume that the top quark is decoupled (the limit $m_t=\infty$).
The correction arising from the virtual top quark was found
in Ref.\ 28 and for the singlet diagrams in Ref.\ 29.
These corrections
have to be added to the above expression for a precision numerical analyses.
Finally, we note that the similar results exist for the pseudoscalar
(MSSM) Higgs boson\cite{MSSM}.  We refer the readers to the original
papers and review articles for the issues that we were not able to
discuss here.

We want to thank the organizers, especially J.~Tr\^an Thanh V\^an
for invitation and the excellent organization
of this meeting. This work was supported by the US Department of Energy
under grant No. DE-FG06-85ER-40224.

\end{document}